\documentclass[runningheads]{llncs}
\usepackage[T1]{fontenc}
\usepackage{graphicx}
\usepackage{amsmath}
\usepackage{amssymb}
\usepackage{enumerate}
\usepackage{tikz}
\usepackage{threeparttable}
\usepackage{multirow}
\usepackage{colortbl}
\usepackage{tabularx}
\usepackage{booktabs}
\usepackage{diagbox}
\usepackage{nicefrac}
\usepackage{multirow}
\newcolumntype{Y}{>{\centering\arraybackslash}X}
\usepackage[caption=false]{subfig}
\usepackage{hyperref}
\usepackage{algorithm}
\usepackage{algpseudocode}

\usepackage[algo2e,linesnumbered,ruled,lined]{algorithm2e}

\newcommand\speck{\ifmmode\text{\scshape Speck}\else{\scshape Speck}\fi}

\newcommand\simon{\ifmmode\text{\scshape Simon}\else{\scshape Simon}\fi}

\newcommand\simeck{\ifmmode\text{\scshape Simeck}\else{\scshape Simeck}\fi}

\newcommand\etal{\textit{et al.}}
\newcounter{experiment}

\usepackage{bbding}
\usepackage[marginal]{footmisc}

\begin{document}
\title{Improved Differential-neural Cryptanalysis for Round-reduced \simeck32/64
\thanks{Supported by organization x.\\
First Author and Second Author contribute equally to this work.\\}
}

%

 \author{Liu Zhang\inst{1,3}\orcidID{0000-0001-6106-3767} \and
 Jinyu Lu\inst{2(}\Envelope\inst{)}\orcidID{0000-0002-7299-0934} \and \\
Zilong Wang\inst{1,3}\orcidID{0000-0002-1525-3356} \and
 Chao Li\inst{2,3}\orcidID{0000-0001-7467-7573}
}

\institute{School of Cyber Engineering, Xidian University, Xi'an 710126, China \\ \email{\{liuzhang@stu., zlwang@\}xidian.edu.cn}
\and
College of Sciences, National University of Defense Technology, Hunan, Changsha 410073, China,
\email{jinyu\_smile@foxmail.com, lichao\_nudt@sina.com} \and
State Key Laboratory of Cryptology, P.O.Box 5159, Beijing 100878, China
}

\maketitle
\begin{abstract}
In CRYPTO 2019, Gohr presented differential-neural cryptanalysis by building the differential distinguisher with a neural network, achieving practical 11-, and 12-round key recovery attack for \speck32/64.
Inspired by this framework, we develop the Inception neural network that is compatible with the round function of \simeck{} to improve the accuracy of the neural distinguishers, thus improving the accuracy of (9-12)-round neural distinguishers for \simeck32/64.
To provide solid baselines for neural distinguishers, we compute the full distribution of differences induced by one specific input difference up to 13-round \simeck32/64. 
Moreover, the performance of the DDT-based distinguishers in multiple ciphertext pairs is evaluated.
Compared with the DDT-based distinguishers, the 9-, and 10-round neural distinguishers achieve better accuracy.
Also, an in-depth analysis of the wrong key response profile revealed that the 12-th and 13-th bits of the subkey have little effect on the score of the neural distinguisher, thereby accelerating key recovery attacks.
Finally, an enhanced 15-round and the first practical 16-, and 17-round attacks are implemented for \simeck32/64, and the success rate of both the 15-, and 16-round attacks is almost 100\%.
\end{abstract}

\section{Introduction}
Lightweight block ciphers present trade-offs between appropriate security and small resource-constrained devices, which is an essential foundation for data confidentiality in resource-constrained environments.
Therefore, the design requirements and security analysis of lightweight block ciphers are of great importance.
Combining traditional analysis methods with ``machine speed'' to efficiently and intelligently evaluate the security of cryptographic algorithm components,
is one of the critical points and trends of current research.
The development of Artificial Intelligence (AI) provides new opportunities for cryptanalysis.
\par
In CRYPTO 2019~\cite{gohr2019improving}, Gohr creatively combines deep learning with differential cryptanalysis and applies it to the \speck32/64, gaining the neural distinguisher ($\mathcal{ND}$) can surpass the DDT-based distinguisher ($\mathcal{DD}$).
Then, a hybrid distinguisher ($\mathcal{HD}$) consisting of a $\mathcal{ND}$ and a classical differential ($\mathcal{CD}$) with highly selective key search strategies result in forceful practical 11-, and 12-round key recovery attacks.
In EUROCRYPT 2021~\cite{benamira2021deeper}, Benamira \emph{et al.} proposed a thorough analysis of Gohr’s neural network. They discovered that these distinguishers are basing their decisions on the ciphertext pair difference and the internal state difference in penultimate and antepenultimate rounds.
\par
To attack more rounds, the component $\mathcal{CD}$ or $\mathcal{ND}$ must be extended.
In ASIACRYPT 2022 ~\cite{bao2022enhancing}, Bao \emph{et al.} devised the first practical 13-round and an improved 12-round $\mathcal{ND}$-based key recovery attacks for \speck32/64 by enhancing the $\mathcal{CD}$s, which they deeply explored more generalized neutral bits of differentials, \emph{i.e.}, conditional (simultaneous) neutral bit/bit-sets. In addition, they obtained $\mathcal{ND}$s up to 11-round \simon32/64 by using  DenseNet and SENet, thus launching the practical 16-round key recovery attack. Zhang \emph{et al.}~\cite{zhang2022improving} focused on improving the accuracy of $\mathcal{ND}$ and added the Inception composed of the multiple-parallel convolutional layers before the Residual network to capture information on multiple dimensions. Under the combined effect of multiple improvements, they reduced the time complexity of key recovery attacks for 12-, and 13-round \speck32/64 and 16-round \simon32/64. They also devised the first practical 17-round key recovery for \simon32/64.
\par
The \simeck{} algorithm~\cite{yang2015simeck}, which combines the good design components from both \simon{} and \speck~\cite{beaulieu2015simon} designed by National Security Agency (NSA), has received a lot of attention for its security. In 2022, Lyu \emph{et al.}~\cite{lyu2022deep} improved Gohr's framework and applied it to \simeck32/64. They obtained (8-10)-round $\mathcal{ND}$s for \simeck32/64 and successfully accomplished attacks for (13-15)-round \simeck32/64 with low data complexity and time complexity. In the same year, Lu \emph{et al.}~\cite{lu2022improved} adopted the multiple ciphertext pairs (8 ciphertext pairs) to train the SE-ResNet neural network fed with a new data format for \simon{} and \simeck{}. Finally, they obtained (9-12)-round $\mathcal{ND}$s for \simeck32/64. This raises the question of whether the key recovery attack for \simeck{} can be enhanced.
\\
\par
\noindent\textbf{Our Contribution.} \noindent The contributions of this work are summarized as follows.
\begin{itemize}
\item[$\bullet$] We improved the Inception neural network proposed by zhang \etal{}~\cite{zhang2022improving} according to the number of cyclic rotation in the round function of \simeck32/64.
Meanwhile, to capture the connections between ciphertext pairs, we use multiple ciphertext pairs forming a sample as the input of the neural network.
Therefore, we improved the accuracy of (9-12)-round $\mathcal{ND}$s using the basic training method and staged training method.
The result can be seen in Table~\ref{tab:ND result}.

\item[$\bullet$] To provide solid baselines for $\mathcal{ND}$s, the full distribution of differences induced by the input difference $(0x0000, 0x0040)$ is computed up to 13 rounds for \simeck32/64.
Also, to make a fair comparison with $\mathcal{ND}$s, the accuracy of the $\mathcal{DD}$s with multiple ciphertext pairs under independent assumptions is investigated.
The comparison shows that the 9-, and 10-round $\mathcal{ND}$s achieve higher accuracy than the $\mathcal{DD}$s, \emph{i.e.}, the $\mathcal{ND}$ contains more information than the $\mathcal{DD}$s (see Table~\ref{tab:ND result}).

\item[$\bullet$] 
Based on the wrong key random hypothesis, we computed the score of the $\mathcal{ND}$ for ciphertexts decrypted with different wrong keys and derived the wrong key response profile (see Figure~\ref{fig:wrkp}).
Through a thorough study of the wrong key response profile, we found that the 12-th and 13-th bit subkeys have little effect on the score of the $\mathcal{ND}$, but the $\mathcal{ND}$ is extremely sensitive to the 14-th, and 15-th bit subkeys. Thus optimizing the Bayesian key search algorithm (see Algorithm~\ref{BayesianKeySearch}) and accelerating the key recovery attack.

\item[$\bullet$] We enhanced the 15-round and launched the first practical 16-, 17-round key recovery attacks for \simeck32/64 based on the $\mathcal{ND}$.
 Table~\ref{tab:attack_result} provides a summary of these results\footnote {The experiment is conducted by Python 3.7.15 and Tensorflow 2.5.0 in Ubuntu 20.04. The device information is Intel Xeon E5-2680V4*2 with 2.40GHz, 256GB RAM, and NVIDIA RTX3080Ti 12GB*6. The source code is available on GitHub \url{https://github.com/CryptAnalystDesigner/Differential-Neural-Cryptanalysis-Simeck32.git}.}.
\end{itemize}
\vspace{-0.5cm}

\begin{table}
	\centering
	\renewcommand\arraystretch{1.3}
	\caption{Summary of key recovery attacks on \simeck32/64}
	\label{tab:attack_result}
	\setlength{\tabcolsep}{1.5mm}{
	\begin{tabular}{ccccccc}
		\toprule
		Attacks        & $R$      & Configure       & Data      & Time               & Success Rate    &  Ref.         \\ \midrule
		\multirow{6}{*}{$\mathcal{ND}$} &  13    & 1+2+9+1    & $2^{16}$  & $2^{27.95+5^\star}$        & $88\%$         &  \cite{lyu2022deep}   \\ \cline{2-7}
		 &  14    & 1+3+9+1    & $2^{23}$  & $2^{32.99+5^\star}$        & $88\%$         &  \cite{lyu2022deep}      \\ \cline{2-7}
		 &  \multirow{2}{*}{15}    & 1+3+10+1    & $2^{24}$  & $2^{33.90+5^\star}$        & $88\%$         &  \cite{lyu2022deep}      \\
		 &      & 1+3+10+1    & $2^{22}$  & $2^{35.309}$        & $99.17\%$      &  Sect.~\ref{sec5:key recovery}       \\	\cline{2-7}
		 &  16    & 1+3+11+1    & $2^{24}$  & $2^{38.189}$        & $100\%$       &  Sect.~\ref{sec5:key recovery}        \\ \cline{2-7}
		 &  17    & 1+3+12+1    & $2^{26}$  & $2^{45.037}$        & $30\%$        &  Sect.~\ref{sec5:key recovery}   \\ \bottomrule
	\end{tabular}}

\begin{enumerate}
\item  $^\star$: Time complexity is calculated in terms of the number of full rounds of \simeck32/64 encryption per second of $2^{23.304}$ in \cite{lyu2022deep}. For a fair comparison, we convert the time complexity to be calculated in terms of the number of 1-round decryption performed per second. These two benchmarks differ by about $2^5$.
\item  Time complexity is calculated based on that one-second equals to $2^{26.693}$ 1-round decryption per second in this paper.
Also, $2^{21.762}$ full-rounds of \simeck32/64 encryption per second can be performed on our device.
\end{enumerate}
\end{table}
\vspace{-0.5cm}
\noindent\textbf{Organization}. The rest of the paper is organized as follows. Section~\ref{sec2:pre} introduces the design of \simeck{} and gives the preliminary on the $\mathcal{ND}$ model. Section~\ref{sec3:nd} gives the data format, network structure, training method, and result of $\mathcal{ND}$s for \simeck32/64. Section~\ref{sec4: nb and wkrp} describes the neutral bits and wrong key response profiles used for key recovery attacks. Section~\ref{sec5:key recovery} exhibits details of the (15-17)-round key recovery attacks. Section~\ref{sec6:con} concludes this paper.
\section{Preliminary}\label{sec2:pre}
In this paper, we denote an $n$-bit binary vector by $x=(x_{n-1},\ldots,x_0)$, where $x_i$ is the  bit in position $i$ with $x_0$ the least significant one. $\oplus$ and $\odot$ denote the eXclusive-OR operation and the bitwise AND operation, respectively. $x \lll \gamma$ or $S^{\gamma}(x)$ represent circular left shift of $x$ by $\gamma$ bits. $x\ggg \gamma$ or $S^{-\gamma}(x)$ represent circular right shift of $x$ by $\gamma$ bits. $x \ \Vert \ y$ represents the concatenation of  bit strings $x$ and $y$.
\subsection{A Brief Description of \simeck}
The \simeck{} family of lightweight block cipher was designed by Yang~\etal{} in CHES 2015~\cite{yang2015simeck}. 
To develop even more compact and efficient block ciphers, it incorporates good design components from both \simon{} and \speck{} designed by NSA. 
A standardized approach for lightweight cryptography was proposed by the National Institute of Standards and Technology (NIST) in 2019. 
Some ideas for this project use modified \simeck{} as a fundamental module, such as ACE~\cite{aagaard2019ace}, SPOC~\cite{altawy2019spoc}, and SPIX~\cite{altawy2019spix}, which suggests that \simeck{} has more practical promise.
\par

\simeck{} adopt the feistel structure to perform encryptions or decryptions on $2n$-bit message blocks using a $4n$-bit key, while $n$ is the word size. The round function of \simeck{} is defined as $f_{5,0,1}(x) = \left(S^5\left(x\right) \odot x \right) \oplus S^1(x)$. Designers reuse the round function in the key schedule to subkeys like \speck{} does. The encryption algorithm of \simeck32/64 is listed in Algorithms~\ref{Encryption of simeck}.

\begin{algorithm2e}[H]
		
		\caption{Encryption of \simeck32/64.}
		\label{Encryption of simeck}
		\LinesNumbered
		\KwIn{$P=(x_0, y_0)$: the paintext, $(k_0, k_1, \cdots, k_{31})$: the round keys.}
		\KwOut{$C=(x_{32}, y_{32})$: the ciphertext.}

		\For{ $r=0 \text{ to }31$}{
			    $x_{r+1} \leftarrow  (x_r \lll 5) \ \& \  x_r \oplus (x_r \lll 1)$\\
                $y_{r+1} \leftarrow  x_r $\\
		} 
\end{algorithm2e}

\subsection{Overview of Neural Distinguisher Model}
The $\mathcal{ND}$ is a supervised model which distinguishes whether ciphertexts are encrypted by plaintexts that satisfies a specific input difference or by random numbers.
Given $m$ plaintext pairs $\{(P_{i, 0},P_{i, 1})\allowbreak, i \in[0, m-1]\}$ and target cipher, the resulting ciphertext pairs $\{(C_{i, 0},C_{i, 1}), i \in[0, m-1]\}$ is regarded as a sample.
Each sample will be attached with a label $Y$:
\[Y=\left\{\begin{array}{l}
    1, \text { if } P_{i, 0} \oplus P_{i, 1}=\Delta,\ i \in[0, m-1] \\
    0, \text { if } P_{i, 0} \oplus P_{i, 1} \neq \Delta,\ i \in[0, m-1]
\end{array}\right.\]
A large number of samples are fed into the neural network for training.
Then, the $\mathcal{ND}$ model can be described as:
\[\begin{array}{c}
    \operatorname{Pr}(Y=1 \mid X_{0}, \ldots, X_{m-1})=F\left(f(X_{0}), \cdots, f(X_{m-1}), \varphi(f(X_{0}), \cdots, f(X_{m-1}))\right), \\
    X_{i}=(C_{i, 0}, C_{i, 1}), i \in[0, m-1], \\
    \operatorname{Pr}(Y=1 \mid X_{0}, \cdots, X_{m-1}) \in[0,1],
\end{array}\]
where $f(X_i)$ represents the basic features of a ciphertext pair $X_i$, $\varphi(\cdot)$ is the derived features, and $F(\cdot)$ is the new posterior probability estimation function.
\section{Neural Distinguisher for \simeck32/64}
\label{sec3:nd}
It is crucial that a well-performing $\mathcal{ND}$ be obtained before a key recovery can be conducted. In this section, we provided the state-of-the-art $\mathcal{ND}$s for \simeck32/64. More importantly, the $\mathcal{DD}$s resulting from the input difference $(0x0000, 0x0040)$ are computed up to 13 rounds for \simeck{}32/64. These $\mathcal{DD}$s provide a solid baseline for $\mathcal{ND}$s. 

\subsection{Construction of the Dataset}
Data quality is fundamentally the most important factor affecting the goodness of a model. Constructing a good dataset for $\mathcal{ND}$s requires answering the following questions:
\begin{enumerate}
    \item[$\bullet$] How to select a good input difference?
    \item[$\bullet$] What data format is used for a sample?
    \item[$\bullet$] How many ciphertext pairs are contained in a sample?
\end{enumerate}
\par
\noindent \textbf{Input Difference.} Numerous experiments have shown that the input difference has a significant impact on the accuracy of the $\mathcal{ND}$s/$\mathcal{DD}$s~\cite{bao2022enhancing,bellini2022cipher,benamira2021deeper,gohr2019improving,gohr2022assessment,hou2021improve,lyu2022deep,yadav2021differential}. Simultaneously, obtaining better results for the key recovery attack depends on whether the input difference of the $\mathcal{ND}$s leads to better accuracy, while leading to the prepended $\mathcal{CD}$s with high probability. Therefore, it is also necessary to consider the number of rounds and the neutral bits of the prepended $\mathcal{CD}$s.
\par
The choice of input difference of $\mathcal{ND}$s varies depending on the block cipher. For \simeck32/64, Lyu \emph{et al.}~\cite{lyu2022deep} present two methods to select the input difference of the $\mathcal{ND}$s. In the first method, the input difference for the $\mathcal{ND}$s is selected from the input difference of the classical differential trail of existing literature. As part of the second method, the MILP model was used to find input differences for classical differential transitions that had high probabilities, then $\mathcal{ND}$s based on these input differences were trained with short epochs, and then the $\mathcal{ND}$s whose input differences had higher accuracy were selected for training long epochs. But they did not consider the effect of the Hamming weight of the input difference on the neural network. Lu \emph{et al.}~\cite{lu2022improved} studied the effect of the input difference of $\mathcal{ND}$s of Hamming weight less than or equal to 3 on the performance of $\mathcal{HD}$s, and their experiments showed that the input difference $(0,e_i)$ is a good choice to obtain a $\mathcal{HD}$ for \simon-like ciphers. Eventually, they built $\mathcal{ND}$s for \simeck32/64 up to 12 rounds with input difference $(0x0000,0x0040)$. 
\par
In this paper, we further explore the neutral bit of the input difference $(0x0000,0x0040)$ (see Sect.~\ref{sec: neutral bit}) and, in a comprehensive comparison, chose this input difference.
\\
\par
\noindent \textbf{Data Format.} In the process of training a $\mathcal{ND}$, the format of the sample needs to be specified in advance. This format is referred to as the $\mathcal{ND}$'s data format for convenience. The most intuitive data format is the ciphertext pair $(C,C') = (x_{r}, y_{r}, x'_{r}, y'_{r})$, which is used in Gohr's network for \speck32/64 in~\cite{gohr2019improving,gohr2022assessment}. As the research progressed, Benamira \emph{et al.}~\cite{benamira2021deeper} constructed a new data format $(x_{r} \oplus x'_{r}, x_{r} \oplus x'_{r} \oplus y_{r} \oplus y'_{r}, x_{r} \oplus y_{r}, x'_{r} \oplus y'_{r})$ through the output of the first convolution layer of Gohr’s neural network for \speck32/64, where $x_{r} \oplus x'_{r}$ represents the left branch difference of the ciphertext, $x_{r} \oplus x'_{r} \oplus y_{r} \oplus y'_{r}$ represents the right branch difference after decrypting one round of ciphertexts without knowing the $(r-1)$-th subkey according to the round function of \speck, $x_{r} \oplus y_{r}$/$x'_{r} \oplus y'_{r}$ represents the right branch ciphertext $C/C'$ of the penultimate round. It shows that the data format is closely related to the structure of the ciphers.
\par
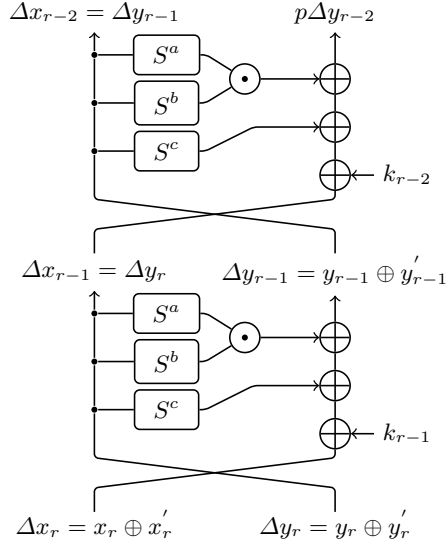
\begin{figure}[htb]
\centering
\begin{tikzpicture}
  [line width=0.6,trim left,
   shiftbox/.style = {
     draw, rounded corners=2pt,
     inner xsep=0.25cm, inner ysep=0.15cm,
   },
   wire/.style = {
     rounded corners=1.5pt
   },
   xor/.style = {
     draw, circle, inner sep=0cm, minimum size=0.4cm,
     append after command = {
       [shorten >=\pgflinewidth, shorten <=\pgflinewidth,]
       (\tikzlastnode.north) edge (\tikzlastnode.south)
       (\tikzlastnode.east) edge (\tikzlastnode.west)
     }
   },
   odot/.style = {
     draw, circle, inner sep=0cm, minimum size=0.4cm
   },
   dot/.style = {
     fill, circle, inner sep=0cm, minimum size=0.08cm
   },
   scale=0.8
   ]

  \node at (3,7.7) (xin) {$\Delta x_{r-1}=\Delta y_{r}$};
  \node at (7,7.7) (yin) {$\Delta y_{r-1}=y_{r-1}\oplus y_{r-1}^{'}$};
  \node[dot] at (3,7) (d1) {};
  \node[dot] at (3,6.2) (d2) {};
  \node[dot] at (3,5.4) (d3) {};
  \node[shiftbox] at (4.2,7) (S1) {$S^a$};
  \node[shiftbox] at (4.2,6.2) (S2) {$S^b$};
  \node[shiftbox] at (4.2,5.4) (S3) {$S^c$};
  \node[xor] at (7,6.6) (x1) {};
  \node[xor] at (7,5.8) (x2) {};
  \node[xor] at (7,5.0) (x3) {};
  \node[odot] at (5.5,6.6) (AND) {};
  \node[dot]  at (5.5,6.6) {};
  \node at (8.2,5.0) (k) {$k_{r-1}$};
  \node at (3, 3.5) () {$\Delta x_r=x_r\oplus x_{r}^{'}$};
  \node at (7, 3.5) () {$\Delta y_r=y_r\oplus y_{r}^{'}$};

  \draw[wire] (d1) -- (S1)  (S1.east) -- +(0.1,0) -- (AND);
  \draw[wire] (d2) -- (S2)  (S2.east) -- +(0.1,0) -- (AND);
  \draw[wire,->] (AND) -- (x1);
  \draw[wire,->] (d3) -- (S3) (S3.east) -- +(0.1,0) -- ++(0.9,0.4) -- (x2);
  \draw[wire,->] (7,3.7) -- (7,4.1) -- (3,4.6) -- (d3) -- (d2) -- (d1) -- (xin) ;
  \draw[wire,->] (3,3.7) -- (3,4.1)-- (7,4.6) -- (x3)-- (x2) -- (x1)  -- (yin);
  \draw[wire,->] (k.west) -- (x3);

  \node at (3,12) (xin) {$\Delta x_{r-2}=\Delta y_{r-1}$};
  \node at (7,12) (yin) {$p\Delta y_{r-2}$};
  \node[dot] at (3,11.3) (d1) {};
  \node[dot] at (3,10.5) (d2) {};
  \node[dot] at (3,9.7) (d3) {};
  \node[shiftbox] at (4.2,11.3) (S1) {$S^a$};
  \node[shiftbox] at (4.2,10.5) (S2) {$S^b$};
  \node[shiftbox] at (4.2,9.7) (S3) {$S^c$};
  \node[xor] at (7,10.9) (x1) {};
  \node[xor] at (7,10.1) (x2) {};
  \node[xor] at (7,9.3) (x3) {};
  \node[odot] at (5.5,10.9) (AND) {};
  \node[dot]  at (5.5,10.9) {};
  \node at (8.2,9.3) (k) {$k_{r-2}$};

  \draw[wire] (d1) -- (S1)  (S1.east) -- +(0.1,0) -- (AND);
  \draw[wire] (d2) -- (S2)  (S2.east) -- +(0.1,0) -- (AND);
  \draw[wire,->] (AND) -- (x1);
  \draw[wire,->] (d3) -- (S3) (S3.east) -- +(0.1,0) -- ++(0.9,0.4) -- (x2);
  \draw[wire,->] (7,8)-- (7,8.4)-- (3,8.9)-- (d3)-- (d2)-- (d1)-- (xin);
  \draw[wire,->] (3,8)-- (3,8.4)-- (7,8.9)-- (x3)-- (x2) -- (x1)  --(yin);
  \draw[wire,->] (k.west) -- (x3);

\end{tikzpicture} 
\caption{Notation of the data format for \simon-like ciphers, where $y_{r-1}=S^{a}(y_{r})\odot S^{b}(y_{r})\oplus S^{c}(y_{r})\oplus x_{r}\oplus k_{r-1}\triangleq A \oplus k_{r-1}$,  $y_{r-1}^{'}=S^{a}(y_{r}^{'})\odot S^{b}(y_{r}^{'})\oplus S^{c}(y_{r}^{'})\oplus x_{r}^{'}\oplus k_{r-1}\triangleq A^{'} \oplus k_{r-1}$, and $p\Delta y_{r-2}=S^{a}(A)\odot S^{b}(A)\oplus S^{c}(A)\oplus y_{r}\oplus S^{a}(A^{'})\odot S^{b}(A^{'})\oplus S^{c}(A^{'})\oplus y_{r}^{'}$}
\label{fig:data-format}
\end{figure}

Bao \emph{et al.}~\cite{bao2022enhancing} accepted data of the form $(x_{r-1},x'_{r-1}, y_{r-1} \oplus y'_{r-1})$ for \simon32/64. Since when the output of the $r$-th round  $(C,C')=(x_{r}, y_{r}, x'_{r}, y'_{r})$ is known, one can directly compute $(x_{r-1},x'_{r-1}, y_{r-1} \oplus y'_{r-1})$ without knowing the $(r-1)$-th subkey according to the round function of \simon-like ciphers. Lu \emph{et al.}~\cite{lu2022improved} further proposed a new data format $(\Delta x_{r},\Delta y_{r}, x_{r}, y_{r}, x'_{r}, y'_{r}, \Delta y_{r-1}, p\Delta y_{r-2})$ and obtained better performance. The details are illustrated in Fig.~\ref{fig:data-format}, and this data format is used in this paper due to its superiority.
\par
\noindent \textbf{Using Multiple Ciphertext Pairs.} Gohr \emph{et al.}~\cite{gohr2022assessment} showed that for a single ciphertext pair, only their differences may provide information for \simon. One option to surpass $\mathcal{DD}$s is to use multiple ciphertext pairs simultaneously, using dependencies between the pairs, especially if the key is fixed. Therefore, in order to surpass $\mathcal{DD}$s, we use multiple ciphertext pairs for training, and the results (Section~\ref{sec3:nd}) confirm that multiple ciphertext pairs indeed help to surpass $\mathcal{DD}$s, albeit only in some rounds. One current trend in deep learning-assisted cryptanalysis is the employment of multiple ciphertext pairs per sample, and our results offer solid evidence in favor of this trend.
\par
The three questions above have been addressed, and the dataset can be generated. Specifically, training and test sets were generated by using the Linux random number generator to obtain uniformly distributed keys $K_{i}$ and multiple plaintext pairs $\{(P_{i,j, 0},P_{i,j, 1})\allowbreak, j \in[0, m-1]\}$ with the input difference $(0x0000,0x0040)$ as well as a vector of binary-valued labels $Y_{i}$.
During the production of the training or test sets for $r$-round \simeck32/64, the multiple plaintext pairs were then encrypted for $r$ rounds if $Y_{i}=1$, while otherwise, the second plaintext of the pairs were replaced with a freshly generated random plaintext and then encrypted for $r$ rounds. Then use the $r$-round ciphertext pairs to generate samples with data of form $(\Delta x_{r},\Delta y_{r}, x_{r}, y_{r}, x'_{r}, y'_{r}, \Delta y_{r-1}, p\Delta y_{r-2})$.

\subsection{Network Architecture}
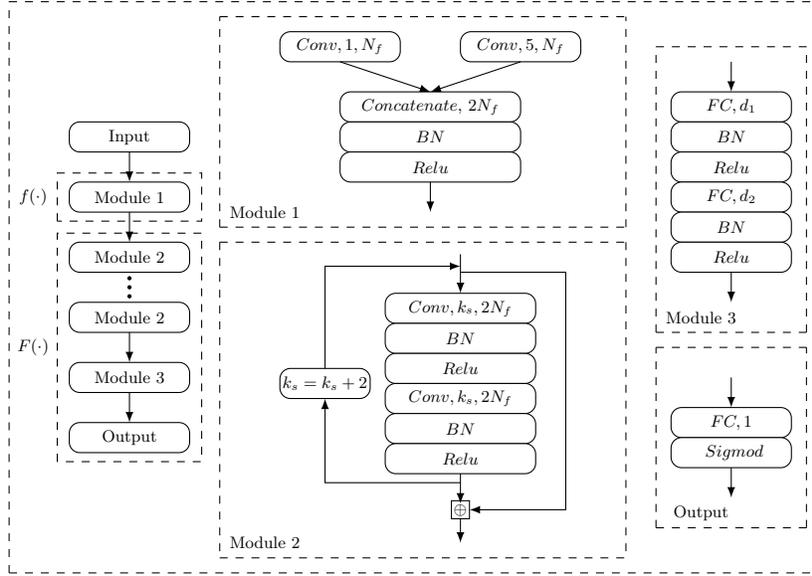
\begin{figure}[htbp]
\centering
\begin{tikzpicture}[help lines/.style={blue!20,very thin} ,scale=0.8, every node/.style={scale=0.7}]

    \draw[dashed] (0,10) -- (13.5,10);
    \draw[dashed] (0,0.5) -- (0,10);
    \draw[dashed] (0,0.5) -- (13.5,0.5);
    \draw[dashed] (13.5,0.5) -- (13.5,10);

    \filldraw[fill=white!20,draw=black,rounded corners] (1,2.5) rectangle (3,3);
    \node at (2,2.75) {Output};
    \filldraw[fill=white!20,draw=black,rounded corners] (1,3.5) rectangle (3,4);
    \node at (2,3.75) {Module 3};\draw[-latex,thin](2,3.5)--(2,3);

    \filldraw[fill=white!20,draw=black,rounded corners] (1,4.5) rectangle (3,5);
    \node at (2,4.75) {Module 2};\draw[-latex,thin](2,4.5)--(2,4);
    \fill (2,5.1) circle[radius=1pt];\fill (2,5.25) circle[radius=1pt];\fill (2,5.4) circle[radius=1pt];
    \filldraw[fill=white!20,draw=black,rounded corners] (1,5.5) rectangle (3,6);
    \node at (2,5.75) {Module 2};

    \filldraw[fill=white!20,draw=black,rounded corners] (1,6.5) rectangle (3,7);
    \node at (2,6.75) {Module 1};\draw[-latex,thin](2,6.5)--(2,6);
    \filldraw[fill=white!20,draw=black,rounded corners] (1,7.5) rectangle (3,8);
    \node at (2,7.75) {Input};\draw[-latex,thin](2,7.5)--(2,7);

    \draw [dashed](0.8,2.35) rectangle(3.2,6.15);
    \draw [dashed](0.8,6.35) rectangle(3.2,7.15);
    \node at (0.4,4.25) {$F(\cdot)$};\node at (0.4,6.75) {$f(\cdot)$};


    \draw[dashed] (3.5,6.25) -- (3.5,9.75);
    \draw[dashed] (3.5,9.75) -- (10.25,9.75);
    \draw[dashed] (3.5,6.25) -- (10.25,6.25);
    \draw[dashed] (10.25,6.25) -- (10.25,9.75);
    \node at (4.25,6.5) {\small{Module 1}};
    \filldraw[fill=white!20,draw=black,rounded corners] (4.5,9) rectangle (6.5,9.5); \node at (5.5,9.25) {$Conv, 1, N_f$};
    \filldraw[fill=white!20,draw=black,rounded corners] (7.5,9) rectangle (9.5,9.5); \node at (8.5,9.25) {$Conv, 5, N_f$};

    \draw[-latex, thin] (5.5,9)--(7,8.5);
    \draw[-latex, thin] (8.5,9)--(7,8.5);

    \filldraw[fill=white!20,draw=black,rounded corners] (5.5,8) rectangle (8.5,8.5); \node at (7,8.25) {$Concatenate$, $2N_f$};
    \filldraw[fill=white!20,draw=black,rounded corners] (5.5,7.5) rectangle (8.5,8); \node at (7,7.75) {$BN$};
    \filldraw[fill=white!20,draw=black,rounded corners] (5.5,7) rectangle (8.5,7.5); \node at (7,7.25) {$Relu$};
    \draw[-latex, thin](7,7)--(7,6.5);

    \draw[dashed] (3.5,6) -- (3.5,0.75);
    \draw[dashed] (3.5,0.75) -- (10.25,0.75);
    \draw[dashed] (3.5,6) -- (10.25,6);
    \draw[dashed] (10.25,6) -- (10.25,0.75);
    \node at (4.25,1) {\small{Module 2}};

    \filldraw[fill=white!20,draw=black,rounded corners] (4.5,3.4) rectangle (6,3.9); \node at (5.25,3.65) {$k_{s}=k_{s}+2$};
    \filldraw[fill=white!20,draw=black,rounded corners] (6.25,4.65) rectangle (8.75,5.15); \node at (7.5,4.9) {$Conv,k_s, 2N_f$};
    \filldraw[fill=white!20,draw=black,rounded corners] (6.25,4.15) rectangle (8.75,4.65); \node at (7.5,4.4) {$BN$};
    \filldraw[fill=white!20,draw=black,rounded corners] (6.25,3.65) rectangle (8.75,4.15); \node at (7.5,3.9) {$Relu$};
    \draw[-latex, thin](7.5,5.8)--(7.5,5.15);

    \filldraw[fill=white!20,draw=black,rounded corners] (6.25,3.15) rectangle (8.75,3.65); \node at (7.5,3.4) {$Conv, k_s, 2N_f$};
    \filldraw[fill=white!20,draw=black,rounded corners] (6.25,2.65) rectangle (8.75,3.15); \node at (7.5,2.9) {$BN$};
    \filldraw[fill=white!20,draw=black,rounded corners] (6.25,2.15) rectangle (8.75,2.65); \node at (7.5,2.4) {$Relu$};
    \draw[-latex, thin](7.5,2.15)--(7.5,1.65);

    \draw (7.35,1.4) rectangle (7.65,1.7);
    \node at (7.5,1.55) {$\oplus$};
    \draw[-latex,thin](7.5,1.4)--(7.5,1);
    \draw[-latex,thin] (7.5,5.5)--(9.25,5.5)--(9.25,1.55)--(7.65,1.55);

    \draw[-latex,thin] (5.25,3.9)--(5.25,5.6)--(7.5,5.6);
    \draw[-latex,thin] (7.5,2)--(5.25,2)--(5.25,3.4);


    \draw[dashed] (10.75,4.5) -- (10.75,9.25);
    \draw[dashed] (10.75,4.5) -- (13.25,4.5);
    \draw[dashed] (13.25,4.5) -- (13.25,9.25);
    \draw[dashed] (10.75,9.25) -- (13.25,9.25);
    \node at (11.5,4.75) {\small{Module 3}};
    
    \draw[-latex, thin](12,9)--(12,8.5);
    \filldraw[fill=white!20,draw=black,rounded corners] (11,8) rectangle (13,8.5);    \node at (12,8.25) {$FC,d_1$};
    \filldraw[fill=white!20,draw=black,rounded corners] (11,7.5) rectangle (13,8);    \node at (12,7.75) {$BN$};
    \filldraw[fill=white!20,draw=black,rounded corners] (11,7) rectangle (13,7.5);    \node at (12,7.25) {$Relu$};
    \filldraw[fill=white!20,draw=black,rounded corners] (11,6.5) rectangle (13,7);    \node at (12,6.75) {$FC,d_2$};
    \filldraw[fill=white!20,draw=black,rounded corners] (11,6) rectangle (13,6.5);    \node at (12,6.25) {$BN$};
    \filldraw[fill=white!20,draw=black,rounded corners] (11,5.5) rectangle (13,6);    \node at (12,5.75) {$Relu$};
    \draw[-latex, thin](12,5.5)--(12,5);

    \draw[dashed] (10.75,1.25) -- (10.75,4.25);
    \draw[dashed] (10.75,4.25) -- (13.25,4.25);
    \draw[dashed] (10.75,1.25) -- (13.25,1.25);
    \draw[dashed] (13.25,1.25) -- (13.25,4.25);
    \node at (11.5,1.5) {\small{Output}};
    \filldraw[fill=white!20,draw=black,rounded corners] (11,2.75) rectangle (13,3.25); \node at (12,3) {$FC,1$};
    \filldraw[fill=white!20,draw=black,rounded corners] (11,2.25) rectangle (13,2.75); \node at (12.05,2.5) {$Sigmod$};
    \draw[-latex, thin](12,3.75)--(12,3.25);
    \draw[-latex, thin](12,2.25)--(12,1.75);

\end{tikzpicture}
\caption{The network architecture for \simeck32/64}
\label{fig:speck_model}
\end{figure}

In CRYPTO 2019, Gohr~\cite{gohr2019improving} used the Residual Network to capture the differential information between the ciphertext pairs, thus getting the $\mathcal{ND}$ for \speck32/64.
To learn the XOR relation at the same position of the ciphertext, a one-dimensional convolution of kernel size 1 is used in Gohr's network architecture.
Since there may be some intrinsic connection between several adjacent bits, Zhang \etal{}~\cite{zhang2022improving} added multiple one-dimensional convolutional layers with different kernel sizes in front of the residual block according to the circular shift operation in the round function of \speck32/64 and \simon32/64.
In this paper, we improved Zhang \etal{}'s neural network to fit with the round function of \simeck{} to improve the accuracy of the $\mathcal{ND}$s, the framework shown in Fig.~\ref{fig:speck_model}.
\\
\par
\noindent \textbf{Initial Convolution (Module 1).} The input layer is connected to the initial convolutional layer, which comprises two convolutional layers with $N_f$ channels of kernel sizes 1 and 5.
The two convolution layers are concatenated at the channel dimension.
Batch normalization is applied to the output of the concatenate layers.
Finally, rectifier nonlinearity is applied to the output of batch normalization, and the resulting $[m,\omega,2N_{f}]$ matrix is passed to the convolutional blocks layer where $m=8$, $\omega=16$ and $N_{f}=32$.
\\
\par
\noindent \textbf{Convolutional Blocks (Module 2).} Each convolutional block consists of two layers of $2N_{f}$ filters.
Each block applies first the convolution with kernel size $k_s$, then a batch normalization, and finally a rectifier layer.
At the end of the convolutional block, a skip connection is added to the output of the final rectifier layer of the block to the input of the convolutional block. It transfers the result to the next block.
After each convolutional block, the kernel size $k_s$ increases by 2 where $k_s=3$.
The number of convolutional blocks is 5 in our model.
\\
\par
\noindent \textbf{Prediction Head (Module 3 and Output).} The prediction head consists of two hidden layers and one output unit.
The three fully connected layers comprise $d_1$, $d_2$ units, followed by the batch normalization and rectifier layers where $d_1=512$ and $d_2=64$.
The final layer consists of a single output unit using the $Sigmoid$ activation function.

\subsection{The Training method of Differential-Neural Distinguisher}
The accuracy is the most critical indicator reflecting the performance of the neural distinguisher. The following training method was carried out to verify the performance of our $\mathcal{ND}$s.
\\
\par
\noindent \textbf{Basic Training Scheme.} We run the training for 20 epochs on the dataset for $N=2*10^7$ and $M=2*10^6$.
We set the batch size to 30000 and used MirroredStrategy of TensorFlow to distribute it equally among the 6 GPUs.
Optimization was performed against mean square error loss plus a small penalty based on L2 weights regularization parameter $c=10^{-5}$ using the Adam algorithm \cite{kingma2014adam}.
A cyclic learning rate schedule was applied, setting the learning rate $l_{i}$ for epoch $i$ to $l_{i} = \alpha + \frac{(n-i) \mod (n+1)}{n}\cdot(\beta-\alpha)$ with $\alpha = 10^{-4},\beta = 2 \times 10^{-3}$ and $n=9$.
The networks obtained at the end of each epoch were stored, and the best network by validation loss was evaluated against a test set.
\\
\par
\noindent \textbf{Training using the Staged Train Method.}
We use several stages of pre-training to train an r-round $\mathcal{ND}$ for \simeck.
First, we use our $(r-1)$-round distinguisher to recognize $(r-3)$-round \simeck{} with the input difference $(0x0140,\allowbreak 0x0080)$ (the most likely difference to appear three rounds after the input difference $(0x0000,0x0040)$.
The training was done on $2*10^7$ instances for 10 epochs with a cyclic learning rate schedule $(2 \times 10^{-3},10^{-4})$.
Then we trained the distinguisher to recognize r-round \simeck{} with the input difference $(0x0000,0x0040)$ by processing $2*10^7$ freshly generated instances for 10 epochs with a cyclic learning rate schedule $( 10^{-4},10^{-5})$.
Finally, the learning rate was dropped to $10^{-5}$ after processing another $2*10^7$ new instances for 10 epochs.

\subsection{Compared Result}
\label{subsec: ND result}
We presented the state-of-the-art $\mathcal{ND}$s for \simeck32/ \allowbreak64.
Meanwhile, we calculate the $\mathcal{DD}$s for \simeck32/64 triggered by the input difference $(0x0000, 0x0040)$ up to 13 rounds to give baselines for $\mathcal{ND}$s (see Table~\ref{tab:DDT combine}).
This is accomplished through the use of the frameworks of Gohr's implementation for \speck32/64 and Bao \emph{et al.}'s implementation for \simon32/64.
The calculation is feasible on \simeck32/64 but quite expensive. In fact, the calculation took about 939 core-days of computation time and yielded about 34 gigabytes of distribution data for each round, which was saved on disk for further studies.
\par
%
%
%
%
%
%
%
%
%
%
%
%
%

\begin{table}[htbp]
	\centering
	\renewcommand\arraystretch{1.3}
	\caption{Accuracy of the $\mathcal{DD}$s for \simeck32/64 with input difference $(0x0000,0x0040)$. Combined means that the corresponding single pair distinguisher was used by combining the scores under independence assumption. For this, $2\times10^6$ samples, each consisting of the given number of pairs $m$, were used to evaluating the accuracy.}
	\label{tab:DDT combine}
	\setlength{\tabcolsep}{1.4mm}{
	\begin{tabular}{c|ccccccccc}
		\toprule
		\diagbox{$R$}{$m$}  &1     &2      & 4    &8        &16      &32   &64    &128    &256 \\	\midrule
		7  &0.9040 &0.9765 & 0.9936 & 0.9996 & 1.0 & 1.0 & 1.0 &1.0 &1.0 \\	\rowcolor{gray!20}
		8  & 0.7105 &0.7921 & 0.8786 & 0.9518 & 0.9907 & 0.9995 & 1.0 &1.0 &1.0 \\
		9  & 0.5738 &0.6097 & 0.6590 & 0.7221 & 0.8011 & 0.8848 & 0.9554 &0.9919 & 0.9998 \\ \rowcolor{gray!20}
		10	&0.5194	&0.5299	&0.5462	&0.5677	&0.5984	&0.6403	&0.6977	&0.7690	&0.8517	\\
		11	&0.5044	&0.5068	&0.5109	&0.5176	&0.5247	&0.5364	&0.5530	&0.5761	&0.6085	\\	\rowcolor{gray!20}
		12	&0.5010	&0.5017	&0.5025	&0.5039	&0.5055	&0.5083	&0.5121	&0.5176	&0.5259	\\
		13	&0.5002	&0.5001	&0.5007	&0.5009	&0.5012	&0.5016	&0.5032	&0.5039	&0.5086	\\ \bottomrule
	\end{tabular}}
\end{table}

It is important to note that when multiple ciphertext pairs are used as a sample in the $\mathcal{ND}$s, comparing the accuracy of the $\mathcal{DD}$s computed with a single ciphertext pair as a sample is not fair.
Actually, the accuracy of the $\mathcal{DD}$s with multiple ciphertext pairs per sample can be calculated.
This calculation is implicitly used by Gohr in~\cite{gohr2019improving}, and later Gohr \emph{et al.}~\cite{gohr2022assessment} explicitly proposed rules for combining probabilities/distinguisher responses (see Corollary 2 in~\cite{gohr2022assessment}).
One can use this rule to explicitly convert a distinguisher for one ciphertext pair into one for an arbitrary number of ciphertext pairs.
Algorithm~\ref{DDT combine} gives the pseudo-code for computing this distinguisher, and the results are shown in Table~\ref{tab:DDT combine}.
\par
\begin{algorithm2e}[htbp]

		\caption{Convert the $\mathcal{DD}$ for one ciphertext pair into one for an $m$ number of ciphertext pairs.}
		\label{DDT combine}
		\LinesNumbered
		\KwIn{DDT: the $R$ round DDT table; $N$: the number of samples for single ciphertext pairs; $m$: the combined number of ciphertext pairs for one sample.}
		\KwOut{the combined Acc, TPR, TNR with $m$ ciphertext pairs.\\}
		Y $\leftarrow \{\}$ \\
		\For{ $i = 1 \text{ to } N $}{
			Y$\left[i*m\right]$ $\leftarrow$ random$\left\{ {0,1}\right\}$\\
			\For{ $j = 1 \text{ to } {m-1}$}{Y$\left[i*m-j\right]$ $\leftarrow$ Y$\left[i*m\right]$}
		}
		Randomly generate $N*m$ samples $[x_1,x_2,\cdots,x_{N*m}]$ according to Y\\
		Z $\leftarrow \{\}$ \\
		\For{ $i = 1 \text{ to }  N*m$}{
			Z$\left[i\right]$ $\leftarrow$ DDT$\left[x_i\right]$\\
		}
	    Z $\leftarrow$ Z / (Z+$2^{-32}$)\\
	    Z $\leftarrow$ mean(Z.reshape($N$,$m$), axis=1) \\
	    predict\_Y $\leftarrow \{\}$ \\
	    \For{ $i = 1 \text{ to } N*m $}{
	    	\If{ Z$\left[i\right] > 0.5$ }{
	    		predict\_Y$\left[i\right]$ $\leftarrow$ 1\\
	    		}
	    	\Else{predict\_Y$\left[i\right]$ $\leftarrow$ 0}
	    }
  		calculate Acc, TPR, TNR based on (Y, predict\_Y)\\
  		\Return Acc, TPR, TNR

            \tcc{In our experiments, $N$ takes $2^{20}$ when $m$ no more than $2^{10}$.}
\end{algorithm2e}

In addition, $r$-round $\mathcal{ND}$ should be compared with $(r-1)$-round $\mathcal{DD}$. Since the data fed to $r$-round $\mathcal{ND}$ is the value of the ciphertext, one can directly compute the differences on $(r-1)$-round outputs without knowing the subkey.
The results are represented in Table~\ref{tab:ND result}, which shows that we improved the accuracy of the $\mathcal{ND}$s for \simeck32/64. More importantly, it is able to surpass the accuracy of $\mathcal{DD}$s for 9- and 10-round.
\par

\begin{table}
	\centering
	\renewcommand\arraystretch{1.3}
	\caption{Comparison of $\mathcal{ND}$s on \simeck32/64 with 8 ciphertext pairs as a sample. The input difference of $\mathcal{ND}$/$\mathcal{DD}$ is $(0x0000,0x0040)$.  *: The staged training method is used to train $\mathcal{ND}$.}
	\label{tab:ND result}
	\setlength{\tabcolsep}{2.0mm}{
	\begin{tabular}{lcccccc}
		\toprule
		$R$   & Attack   & Network   &Acc  & TPR    & TNR     & Ref. \\ \midrule
		\multirow{3}{*}{9}      &$\mathcal{DD}$  &DDT   & 0.9518 &   0.9604     &   0.9433     & Sect.~\ref{sec3:nd}      \\
		      &$\mathcal{ND}$  &SE-ResNet      & 0.9952  &   0.9989     &   0.9914      & ~\cite{lu2022improved}      \\
		      &$\mathcal{ND}$  &Inception      & 0.9954  &   0.9986     &   0.9920      & Sect.~\ref{sec3:nd}     \\
		\hline

		\multirow{3}{*}{10}      &$\mathcal{DD}$  &DDT  & 0.7221 &   0.7126     &   0.7316      & Sect.~\ref{sec3:nd}      \\
		      &$\mathcal{ND}$  &SE-ResNet  & 0.7354   &   0.7207    &   0.7501      & ~\cite{lu2022improved}       \\
	         &$\mathcal{ND}$  &Inception  & 0.7371  &   0.7165     &   0.7525      & Sect.~\ref{sec3:nd}       \\
		\hline

		\multirow{4}{*}{11}      &$\mathcal{DD}$  &DDT  & 0.5677 &   0.5416     &   0.5940      & Sect.~\ref{sec3:nd}   \\
		      &$\mathcal{ND}$  &SE-ResNet  & 0.5646   &   0.5356      &  0.5936      & ~\cite{lu2022improved}      \\
		      &$\mathcal{ND}$  &Inception & 0.5657  &   0.5363     &   0.5954      & Sect.~\ref{sec3:nd}      \\
		   &$\mathcal{ND}$  &Inception  & 0.5666$^{\star}$  &   0.5441     &   0.5895      & Sect.~\ref{sec3:nd}      \\
		\hline

		\multirow{3}{*}{12}      &$\mathcal{DD}$  &DDT  & 0.5176 &   0.4737     &   0.5615      & Sect.~\ref{sec3:nd}   \\
		    &$\mathcal{ND}$  &SE-ResNet  & 0.5146$^{\star}$  &   0.4770      &  0.5522     & ~\cite{lu2022improved}      \\
		    &$\mathcal{ND}$  &Inception  & 0.5161$^{\star}$  &   0.4807     &   0.5504      & Sect.~\ref{sec3:nd}      \\ \bottomrule
	\end{tabular}}
\end{table}

\section{Neutral bits and Wrong Key Response Profile}
\label{sec4: nb and wkrp}
In Sect.~\ref{sec3:nd}, we provided the state-of-the-art $\mathcal{ND}$s for \simeck32/64, which use to perform better key recovery attacks in the following section. In~\cite{gohr2019improving}, Gohr provides a framework of $(1+s+r+1)$-round key recovery attack (refer to Appendix~\ref{key_recovery_attack}) consisting of three techniques to increase the success rate and speed up the attacks, where $s$ is the length of the $\mathcal{CD}$, and $r$ is the length of the $\mathcal{ND}$. Here is a description of these techniques.
\\
\par
\noindent \textbf{Neutral Bits.} In the key recovery attack, multiple samples (formed into a ciphertext structure) decrypted by the guessed subkey are predicted using the distinguisher. Then, the multiple scores are combined according to formula 
$v_{k}=\sum_{i=1}^{n_{b}}\nicefrac{Z_{i}^{k}}{1-Z_{i}^{k}}$
as the final score of that guessed subkey to reduce the misjudgment rate of the $\mathcal{ND}$. Since the $\mathcal{CD}$ suspended in front of the $\mathcal{ND}$ are probabilistic, resulting in sample entering the distinguisher not satisfying the same distribution. Multiple samples generated by neutral bits will have the same distribution. Also, the lower the accuracy of the distinguisher, the more neutral bits are needed.\\
\par
 \noindent \textbf{Priority of Ciphertext Structure.} Spending the same amount of computation on every ciphertext structure is inefficient. Gohr used a generic method (automatic exploitation versus exploration tradeoff based on Upper Confidence Bounds) to focus the key search on the most promising ciphertext structures.
The priority score of each ciphertext structure is $s_{i} = \omega_{\max}^{i}+\sqrt{n_{c}}\cdot \sqrt{\log_{2}(j)}/n_{i}$ where denote by $\omega_{\max}^{i}$ the highest distinguisher score, $n_{i}$ the number of previous iterations in which the $i$th ciphertext structure, $j$ the number of the current iteration and $\sqrt{n_{c}}$ the number of ciphertext structures available.
\\
\par
\noindent \textbf{Wrong Key Response Profile.} The key search policy based on Bayesian Optimization drastically reduces the number of trial decryptions. The basic idea of this policy is the wrong key randomization hypothesis.
This hypothesis does not hold when only one round of trial decryption is performed, especially in a lightweight cipher.
The expected response of the $\mathcal{ND}$ upon wrong-key decryption will depend on the bitwise difference between the trial and real keys.
This wrong-key response profile can be captured in a precomputation. Give some trial decryptions, the optimization step then trials to come up with a new set of candidate keys to try. These new candidate keys are chosen to maximize the probability of the observed distinguisher responses.

\subsection{Exploring Neutral Bits}
\label{sec: neutral bit}
To be able to attack more rounds with the $\mathcal{ND}$, the $\mathcal{CD}$ is generally prepended in front of the $\mathcal{ND}$. For the resulting $\mathcal{HD}$ used in the key recovery attack, it is not straightforward to aggregate enough samples of the same distribution fed to the $\mathcal{ND}$ due to the prepended $\mathcal{CD}$. To overcome this problem, Gohr~\cite{gohr2019improving} used the neutral bits of the $\mathcal{CD}$. The more neutral bits there are for the prepended $\mathcal{CD}$, the more samples of the same distribution could be generated for the $\mathcal{ND}$. However, generally, the longer the $\mathcal{CD}$, the fewer the neutral bits. Finding enough neutral bits for prepending a long $\mathcal{CD}$ over a weak $\mathcal{ND}$ becomes a difficult problem for devising a key recovery to cover more rounds. To solve this problem, Bao \etal{} exploited various generalized NBs to make weak $\mathcal{ND}$ usable again. Particularly, they employed conditional simultaneous neutral bit-sets (CSNBS) and switching bits for adjoining differentials (SBfAD), which are essential for achieving efficient 12-round and practical 13-round attacks for \speck32/64.
\par
Thus, the first part of the key recovery attack focuses on finding various types of neutral bits. Given a differential, in order to find the neutral bits, it is generally divided into two steps: firstly, collect enough conforming pairs (correct pairs); secondly, flip the target bits of the conforming pair, or flip all the bits contained in the target set of bits, and check the probability that the new plaintext pair is still the conforming pair.
\\
\par
\noindent \textbf{Finding SNBSs for 3-round Differential}. For the prepended 3-round $\mathcal{CD}$ $(0x0140,0x0200) \rightarrow (0x0000,0x0040)$ on top of the $\mathcal{ND}$s, one can experimentally obtain 14 deterministic NBs and 2 SNBSs (simultaneously complementing up to 4 bits) using an exhaustive search.  Concretely, for the 3-round differential $(0x0140,0x0200) \rightarrow (0x0000,0x0040)$, (simultaneous-) neutral bits and bit-sets are [3], [4], [5], [7], [8], [9], [13], [14], [15], [18], [20], [22], [24], [30], [0, 31], [10, 25].
\\
\par
\noindent \textbf{Finding SNBSs for 4-round Differential}. For the prepended 4-round $\mathcal{CD}$ $(0x0300,0x0440) \rightarrow (0x0000,0x0040)$ on top of the $\mathcal{ND}$s, there are 7 complete NB/SNBS: [2], [4], [6], [8], [14], [9, 24], [9, 10, 25].
Still, the numbers of NBs/SNBSs are not enough for appending a weak neural network distinguisher. Thus, conditional ones were searched using Algorithm 3 in paper~\cite{bao2022enhancing}, and the obtained CSNBSs and their conditions are summarized together in Table~\ref{simeck_4csnb}.

 \begin{table}[H]
    \centering
    \renewcommand\arraystretch{1.2}
    \caption{CSNBS for 4-round Classical Differential $(0x0300,0x0440) \rightarrow (0x0000,0x0040)$ of \simeck32/64}
    \label{simeck_4csnb}
    \setlength{\tabcolsep}{2mm}{
        \begin{tabular}{ll|ll}
            \toprule
            Bit-set  & C.   & Bit-set  & C.    \\ \midrule
            \multicolumn{2}{c}{$x[0, 10]$} & \multicolumn{2}{c}{$x[2, 12]$} \\ \midrule \relax
            
            [21] & 00 & [23] & 00  \\ \relax
            [21, 5] & 10 & [23, 12] & 10 \\ \relax
            [21, 10] & 01 & [23, 7] & 01 \\ \relax
            [21, 10, 5] & 11 & [23, 12, 7] & 11 \\
            \bottomrule
    \end{tabular}}
    \begin{tablenotes}
        \item C.: Condition on $x[i,j]$, e.g., $x[i,j] = 10$ means $x[i]=1$ and $x[j]=0$.
    \end{tablenotes}
\end{table}
\subsection{Wrong Key Response Profile}
\begin{figure}[H]
\centering
\subfloat[$\text{WKRP}_{9}$]{\includegraphics[width=0.48\linewidth]{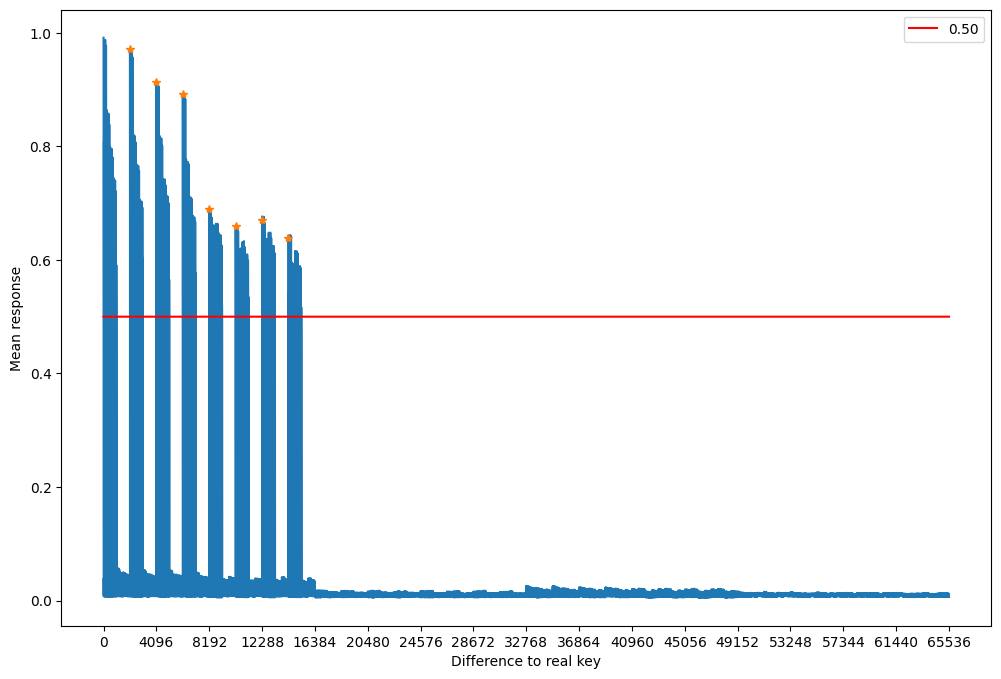}\label{fig:wkrp_9}}
\quad
\subfloat[$\text{WKRP}_{10}$]{\includegraphics[width=0.48\linewidth]{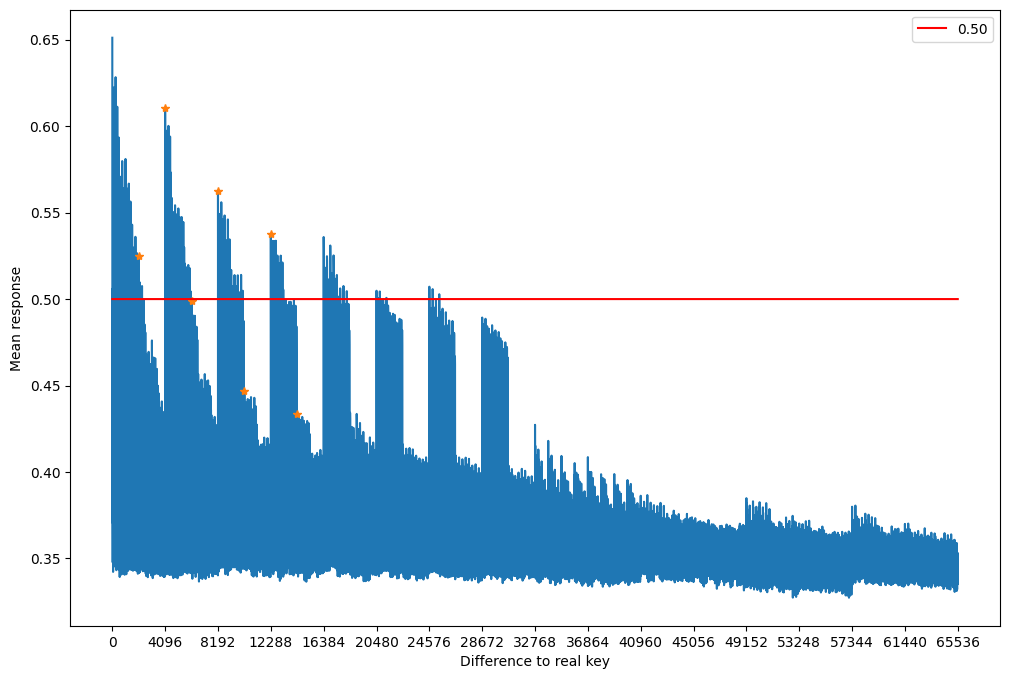}\label{fig:wkrp_10}}
\\
\subfloat[$\text{WKRP}_{11}$]{\includegraphics[width=0.48\linewidth]{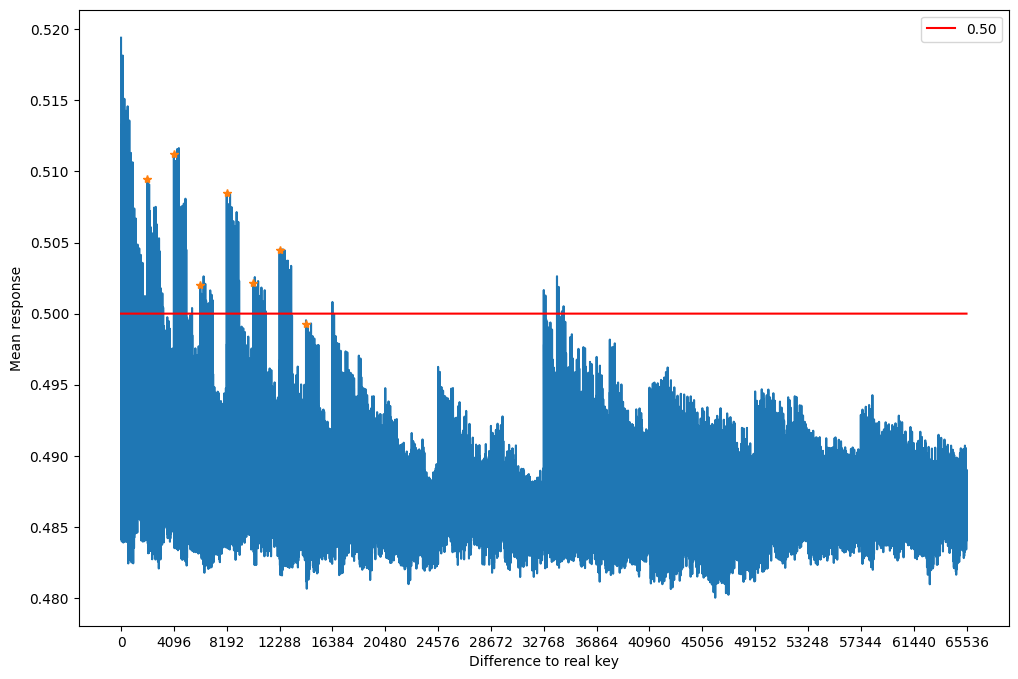}\label{fig:wkrp_11}}
\quad
\subfloat[$\text{WKRP}_{12}$]{\includegraphics[width=0.48\linewidth]{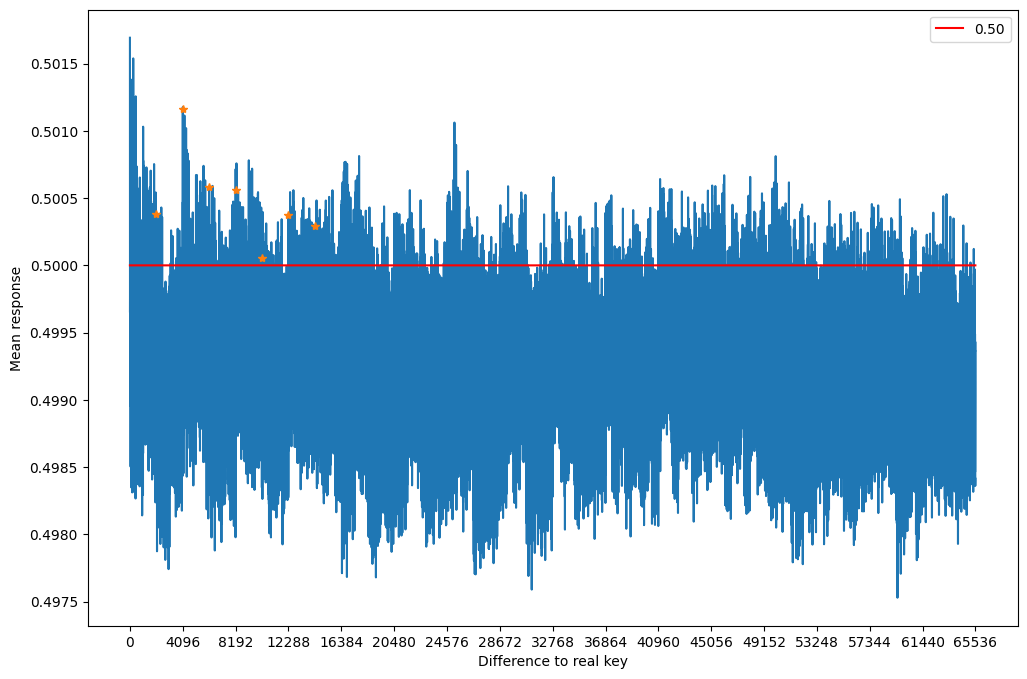}\label{fig:wkrp_12}}
\caption{Wrong Key Response Profile for \simeck32/64.}
\label{fig:wrkp}
\end{figure}

To calculate the $r$-round wrong key response profile, we generated 3000 random keys and multiple input pairs $\{(P_{i,0},P_{i,1}),i\in [0,m-1]\}$ for each difference $\delta \in (0,2^{16})$ and encrypted for $r+1$ rounds to obtain ciphertexts $\{(C_{i,0},C_{i,1}),i\in [0,m-1]\}$, where $P_{i,0}\oplus P_{i,1}=\Delta$. 
Denoting the final real subkey of each encryption operation by $k$, we then performed single-round decryption to get $E_{k\oplus\delta}^{-1}(\{C_{i,0},i\in [0,m-1]\}), E_{k\oplus\delta}^{-1}(\{C_{i,1},i\in [0,m-1]\})$ 
and had the resulting partially decrypted ciphertext pair rated by an $r$-round $\mathcal{ND}$.
$\mu_{\delta}$ and $\sigma_{\delta}$ were then calculated as empirical mean and standard deviation over these 3000 trials.
We call the $r$-round wrong key response profile $\text{WKRP}_r$. From the wrong key Response Profile, we can find some rules to speed up the key recovery attack.
\begin{enumerate}
    \item[$\bullet$] \textbf{Analysis of $\text{WKRP}_{9}$}. In Figure~\ref{fig:wkrp_9}, when the difference between guessed key and real key $\delta$ is greater than 16384, the score of the distinguisher is close to 0. 
    This phenomenon indicates that the score of the distinguisher is very low when the 14-th and 15-th bit is guessed incorrectly.
    When $\delta \in \{2048,4096,8192,10240,12288,\allowbreak14436\}$, the score of the distinguisher is greater than 0.6.
    This indicates that when the 11-th, 12-th, and 13-th bits are guessed incorrectly, it has little effect on the score of the distinguisher. \\
    \item[$\bullet$] \textbf{Analysis of $\text{WKRP}_{10}$ \text{and} $\text{WKRP}_{11}$}. It is clear from Figure~\ref{fig:wkrp_10} that when the $\delta$ is greater than 32768, the score of the distinguisher is less than 0.45, \emph{i.e.}, the 15-th bit has a greater impact on the distinguisher score.
    When $\delta \in \{4096,8192,12288\}$, the score of the distinguisher is close to 0.55.
    This indicates that when the 12-th and 13-th bits are guessed incorrectly, it has little effect on the score of the distinguisher.
    It can also be observed from Figure~\ref{fig:wkrp_11} that the 12-th and 13-th bits have less influence on the score of the distinguisher, and the 14-th and 15-th bits have more influence on the score of the distinguisher.\\
  
    \item[$\bullet$] \textbf{Analysis of $\text{WKRP}_{12}$}. Despite the small difference in scores in Figure~\ref{fig:wkrp_12}, it was found that when only the 12-th and 13-th bits are wrongly guessed, the score of the distinguisher is still higher than the other positions.
\end{enumerate}
\par

From the four wrong key response profiles, we can conclude that when the 14-th and 15-th bit subkeys are guessed incorrectly, it has a greater impact on the score of the distinguisher; when the 12-th and 13-th bit subkeys are guessed incorrectly, it has a smaller impact on the score of the distinguisher. According to these phenomena, we can speed up the key recovery attack.
\begin{enumerate}
    \item [$\bullet$] \textbf{Guess the 14-th and 15-th bit subkeys.} Since the difference between the score of the distinguisher of bits 14 and 15 in the case of correct and incorrect guesses is relatively large, we can first determine the values of these two bits. Before performing a Bayesian key search, a random set of subkeys is guessed, then the 14-th and 15-th bits of the subkeys are traversed, and the ciphertext is decrypted using the subkeys. Thus, the values of the 14-th and 15-th bits can be determined based on the score of the distinguisher.
    The Bayesian key search algorithm can easily recover these two bits even if the values of these two bits are not determined in advance. \\
    \item [$\bullet$] \textbf{Ignore the 12-th and 13-th bit subkeys.} Since the 12-th and 13-th bit subkeys have less influence on the score of the distinguisher, we first set these two bits to 0 when generating the first batch of candidate subkeys and then randomize the values of the two bits after completing the Bayesian key sorting and recommending the new candidate subkeys.
    Previous researchers have also exploited this feature to accelerate key recovery attacks, and the 14-th and 15-th bit subkeys have little impact on the score of the distinguisher when guessed incorrectly for \speck32/64 and \simon32/64\cite{bao2022enhancing,gohr2019improving,zhang2022improving}.
    The Bayesian key search algorithm considering insensitive key bits is shown in Algorithm~\ref{BayesianKeySearch}.
\end{enumerate}

\begin{algorithm2e}[htbp]
		
		\caption{BayesianKeySearch Algorithm For \simeck32/64.}
		\label{BayesianKeySearch}
		\LinesNumbered    
		\KwIn{Ciphertext structure $\mathcal{C}:=\{C_{0},\cdots,C_{n_b-1}\}$, a neural distinguisher $\mathcal{ND}$, and its wrong key response profile $\mu$ and $\sigma$, the number of candidates to be generated within each iteration $n_{cand}$, the number of iterations $n_{byit}$}
		\KwOut{The list $L$ of tuples of recommended keys and their scores}
            $S:= \{k_0,k_1,\cdots,k_{n_{cand}-1}\}\leftarrow$ choose $n_{cand}$ values at random without replacement from the set of all subkey candidates \\
            $S = S \ \& \  \text{0xCFFF} $ \\
            $L \leftarrow \{\}$ \\
		\For{$t=1 \text{ to } n_{byit}$}{
			\For{$\forall k_i \in S$}{
			    \For{$j=0 \text{ to } n_b-1$}{
			             $C_{j,k_i}^{'}=F_{k_i}^{-1}(C_j)$\\
                          $v_{j,k_i}=\mathcal{ND}(C_{j,k_i}^{'})$\\
                          $s_{j,k_i}=\log_2(v_{j,k_i}/(1-v_{j,k_i}))$
		          } 
                    $s_{k_i}=\sum_{j=0}^{n_b-1}s_{j,k_i}$; \tcc{the combined score of $k_i$ using neutral bits.}
                     $L\leftarrow L \| (k_i,s_{k_i})$; \\
                     $m_{k_i}=\sum_{j=0}^{n_b-1}v_{j,k_i}/n_b$ 
		    }
                \For{$k \in \{0,1,\cdots,2^{16}-1\} \ \& \  {\rm 0xCFFF}$}{
                    $\lambda_k=\sum_{i=0}^{n_{cand}-1}(m_{k_i}-\mu_{k_i\oplus k})^{2}/\sigma_{k_i\oplus k}^{2}$; \tcc{ using wrong key response profile.}
                }
                $S\leftarrow {\rm argsort}_{k}(\lambda)[0:n_{cand}-1]$; \\
                $r:=\{r_0,r_1,\cdots,r_{n_{cand}-1}\}\leftarrow$ choose $n_{cand}$ values at $(0,4)$ at random \\ 
                $r = r << 12$; \tcc{ Randomize the 12-th and 13-th bit subkeys.} 
                $S = S \oplus r$ 
		} 
    \Return $L$    
\end{algorithm2e}

\section{Practical Key Recovery Attack}
\label{sec5:key recovery}
When a fast graphics card is used, the performance of the implementation is not limited by the speed of neural network evaluation but by the total number of iterations on the ciphertext structures.
We count a key guess as successful if the sum of the Hamming weights of the
differences between the returned last two subkeys and the real two subkeys are at most two.
The experimental parameters for key recovery attacks are denoted as follows.
\begin{enumerate}[1.]
\item $n_{cts}$: the number of ciphertext structure.
\item $n_{b}$: the number of ciphertext pairs in each ciphertext structures.
\item $n_{it}$: the total number of iterations on the ciphertext structures.
\item $c_1$ and $c_2$: the cutoffs with respect to the scores of the recommended last subkey and second to last subkey, respectively.
\item $n_{byit1}, n_{cand1}$ and $n_{byit2}, n_{cand2}$: the number of iterations and number of key candidates within each iteration in the \textsc{BayesianKeySearch} Algorithm for guessing each of the last and the second to last subkeys, respectively.
\end{enumerate}
\subsection{Complexity Calculation}
\noindent \textbf{Theoretical Data Complexity.} The theoretical data complexity of the experiment is calculated by the formula $n_{b}\times n_{ct} \times m \times 2$.
In the actual experiment, when the accuracy of the $\mathcal{ND}$ is high, the key can be recovered quickly and successfully. Not all the ciphertext structure is used, so the actual data complexity is lower than the theoretical.\\
\par
\noindent \textbf{Experimental Time Complexity.}  
The time complexity calculation formula in our experiments is $2^{26.693}\times rt \times\log_{1-sr}{0.01}$, which is borrowed from~\cite{zhang2022improving}.
Our device can perform $2^{26.693}$ 1-round decryption per second.
$rt$ is the average running time of multiple experiments.
The success rate $sr$ is the number of successfully recovered subkeys divided by the number of experiments.
We calculate how many experiments need to be performed to ensure at least one successful experiment.
When the overall success rate is 99\%, we consider the experiment to be successful, and the number of experiments $ne$ is: $1-(1-sr)^{ne}=0.99$, \emph{i.e.}, $\log_{1-sr}{0.01}$.

\subsection{Key Recovery Attack on 15-round \simeck32/64}
\textbf{Experiment \stepcounter{experiment}\theexperiment:} The components of key recovery attack $\mathcal{A}^{\simeck15R}$ of 15-round \simeck32/64 are as follows.
\begin{enumerate}
    \item 3-round $\mathcal{CD}$ $(0x0140,0x0200)\rightarrow (0x0000,0x0040)$.
    \item neutral bits of generating multiple ciphertext pairs: $[3],[4],[5]$.
    \item neutral bits of combined response of neural distinguisher: $[7],[8],[9],[13],[14],\allowbreak[15],[18],[20]$.
    \item 10-round neural distinguisher $\mathcal{ND}^{\simeck_{10R}}$ and wrong key response profiles $\mathcal{ND}^{\simon_{10R}}\cdot\mu$ and $\mathcal{ND}^{\simeck_{10R}}\cdot\delta$.
    \item 9-round distinguisher $\mathcal{ND}^{\simeck_{9R}}$ and wrong key response profiles $\mathcal{ND}^{\simon_{9R}}\cdot\mu$ and $\mathcal{ND}^{\simeck_{9R}}\cdot\delta$.
\end{enumerate}
Concrete parameters used in our 15-round key recovery attack $\mathcal{A}^{\simeck15R}$ are listed as follows.
\begin{center}
\begin{tabular}{l@{\hspace{0.5cm}}l@{\hspace{0.5cm}}l@{\hspace{0.5cm}}l}
\toprule
$m=8$ & $n_{b} = 2^{8} $  &$n_{cts}= 2^{10}$ & $n_{it}= 2^{11} $  \\
 $c_{1}=10$ & $c_{2} =10$   & $n_{byit1}=n_{byit2}=5$  &  $n_{cand1}=n_{cand2}=32 $  \\ 
 \bottomrule
\end{tabular}
\end{center}
\par
The theoretical data complexity is $ m \times n_{b} \times n_{cts} \times 2= 2^{22}$ plaintexts. 
The actual data complexity is $2^{19.621}$. 
In total, 120 trials are running and 119 successful trials. Thus, the success rate $sr$ is 99.17\%. The average running time of the experiment $rt$ is 407.901s.
The time complexity is $  2^{26.693}\times  rt \times\log_{1-sr}{0.01} = 2^{35.309}$.

\subsection{Key Recovery Attack on 16-round \simeck32/64}
\textbf{Experiment \stepcounter{experiment}\theexperiment:} The components of key recovery attack $\mathcal{A}^{\simeck16R}$ of 16-round \simeck32/64 are shown as follows.
\begin{enumerate}
    \item 3-round $\mathcal{CD}$ $(0x0140,0x0200)\rightarrow (0x0000,0x0040)$.
    \item neutral bits of generating multiple ciphertext pairs: $[3],[4],[5]$.
    \item neutral bits of combined response of neural distinguisher: $[7],[8],[9],[13],[14],\allowbreak[15],[18],[20],[22],[24]$.
    \item 11-round neural distinguisher $\mathcal{ND}^{\simeck_{11R}}$ and wrong key response profiles $\mathcal{ND}^{\simeck_{11R}}\cdot\mu$ and $\mathcal{ND}^{\simeck_{11R}}\cdot\delta$.
    \item 10-round neural distinguisher $\mathcal{ND}^{\simeck_{10R}}$ and wrong key response profiles $\mathcal{ND}^{\simeck_{10R}}\cdot\mu$ and $\mathcal{ND}^{\simeck_{10R}}\cdot\delta$.
\end{enumerate}
Concrete parameters used in our 16-round key recovery attack $\mathcal{A}^{\simeck16R}$ are listed as follows.
\begin{center}
\begin{tabular}{l@{\hspace{0.5cm}}l@{\hspace{0.5cm}}l@{\hspace{0.5cm}}l}
\toprule
$m=8$ & $n_{b} = 2^{10} $  &$n_{cts}= 2^{10}$ & $n_{it}= 2^{11} $  \\
 $c_{1}=10$ & $c_{2} =10$   & $n_{byit1}=n_{byit2}=5$  &  $n_{cand1}=n_{cand2}=32 $  \\
 \bottomrule
\end{tabular}
\end{center}
\par
The theoretical data complexity is $ m\times  n_{b}\times n_{cts}\times 2= 2^{24}$ plaintexts. 
The actual data complexity is $2^{22.788}$.
We use 6 processes, each running 20 experiments. Since the memory limit was exceeded during the experiment, one process was killed, leaving 100 experiments, 100 of which successfully recovered the key.
Thus, the success rate $sr$ is 100\%. The average running time of the experiment $rt$ is 2889.648s.
The time complexity is $  2^{26.693}\times  rt  = 2^{38.189}$.

\subsection{Key Recovery Attack on 17-round \simeck32/64}
\textbf{Experiment \stepcounter{experiment}\theexperiment:} The components of key recovery attack $\mathcal{A}^{\simeck17R}$ of 17-round \simeck32/64 are shown as follows.
\begin{enumerate}
    \item 3-round $\mathcal{CD}$ $(0x0140,0x0200)\rightarrow (0x0000,0x0040)$.
    \item neutral bits of generating multiple ciphertext pairs: $[3],[4],[5]$.
    \item neutral bits of combined response of neural distinguisher: $[7],[8],[9],[13],[14],\allowbreak[15],[18],[20],[22],[24],[30],[0,31]$.
    \item 12-round neural distinguisher $\mathcal{ND}^{\simeck_{12R}}$ and wrong key response profiles $\mathcal{ND}^{\simeck_{12R}}\cdot\mu$ and $\mathcal{ND}^{\simeck_{12R}}\cdot\delta$.
    \item 11-round neural distinguisher $\mathcal{ND}^{\simeck_{11R}}$ and wrong key response profiles $\mathcal{ND}^{\simeck_{11R}}\cdot\mu$ and $\mathcal{ND}^{\simeck_{11R}}\cdot\delta$.
\end{enumerate}
Concrete parameters used in our 17-round key recovery attack $\mathcal{A}^{\simeck17R}$ are listed as follows.
\begin{center}
\begin{tabular}{l@{\hspace{0.5cm}}l@{\hspace{0.5cm}}l@{\hspace{0.5cm}}l}
\toprule
$m=8$ & $n_{b} = 2^{12} $  &$n_{cts}= 2^{10}$ & $n_{it}= 2^{11} $  \\
 $c_{1}=20$ & $c_{2} =-120$   & $n_{byit1}=n_{byit2}=5$  &  $n_{cand1}=n_{cand2}=32$  \\
 \bottomrule
\end{tabular}
\end{center}
\par
The theoretical data complexity is $ m\times n_{b} \times n_{cts} \times 2= 2^{26}$ plaintexts. 
The actual data complexity is $2^{25.935}$. 
In total,  trials are 50 running, and there are 15 successful trials. Thus, the success rate $sr$ is 30\%. The average running time of the experiment $rt$ is 25774.822s.
The time complexity is $  2^{26.693}\times  rt \times\log_{1-sr}{0.01} = 2^{45.037}$.
\begin{remark}
There are two reasons why we do not launch a 17-round key recovery attack using a 4-round $\mathcal{CD}$ and an 11-round $\mathcal{ND}$. 
One is that the probability of the 4-round $\mathcal{CD}$  $(0x0300,0x0440) \rightarrow (0x0000,0x00\allowbreak40)$ is about $2^{12}$ (the probability of the 3-round $\mathcal{CD}$ $(0x0140,0x0200)\rightarrow (0x0000,0x0040)$ is about $2^{-8}$), resulting in too much data required, and the second is that there are not enough neutral bits in the 4-round $\mathcal{CD}$.
\end{remark}

\section{Conclusion}
\label{sec6:con}

In this paper, we show practical key recovery attacks up to 17 rounds of \simeck\allowbreak32/64, raising the technical level of practical attacks by two rounds. We design neural network that fits with the round function of \simeck{} to improve the accuracy of the neural distinguishers, and is able to outperform the DDT-based distinguisher in some rounds. To launch more rounds of the key recovery attack, we make a concerted effort on the classical differential and the neural distinguisher to make both modules good. In addition, we optimize the key recovery attack process by deeply analyzing the wrong key response profile, thus reducing the complexity of the key recovery attack.


\bibliographystyle{splncs04}
\bibliography{ref}
\clearpage
\appendix
\section{Appendix}
\subsection{Procedure of $(1+s+r+1)$-round key recovery attack}
\label{key_recovery_attack}
The attack procedure is as follows.
\begin{enumerate}
\item Initialize variables $Gbest_{key}\leftarrow (\text{None, None})$, $Gbest_{score}\leftarrow -\infty$.
\item Generate $n_{cts}$ random plaintext pairs with difference $\Delta P$.
\item Using $n_{cts}$ plaintext pairs  and $\log_{2}{m}$ neutral bit with probability one to generate $n_{cts}$ multiple plaintext pairs. Every multiple plaintext pairs have $m$ plaintext pairs.
\item From the $n_{cts}$ multiple plaintext pairs, generate $n_{cts}$ plaintext structures using $n_{b}$ generalized neutral bit.
\item Decrypt one round using zero as the subkey for all multiple plaintext pairs in the structures and obtain $n_{cts}$ plaintext structure.
\item Query for the ciphertexts under $(1+s+r+1)$-round \simeck32/64 of the $n_{cts}\times n_{b} \times 2$ plaintext structures, thus obtain $n_{cts}$ ciphertext structures, denoted by $\{\mathcal{C}_{1},\ldots, \mathcal{C}_{n_{cts}}\}$.
\item Initialize an array $\omega_{\text{max}}$ and an array $n_{\text{visit}}$ to record the highest distinguisher score obtained so far and the number of visits have received in the last subkey search for the ciphertext structures.
\item Initialize variables $best_{score} \leftarrow -\infty$, $best_{key} \leftarrow (\text{None, None})$, $best_{pos} \leftarrow \
\text{None}$ to record the best score, the corresponding best recommended values for the two subkeys obtained among all ciphertext structures and the index of this ciphertext structures.
\item For $j$ from 1 to $n_{it}$:
    \begin{enumerate}
    \item Compute the priority of each of the ciphertext structures as follows: $s_{i} = \omega_{\text{max}i}+\alpha\cdot\sqrt{\log_{2}{j}/n_{\text{visit}i}} $, for $i \in \{1,\ldots,n_{cts}\}$, and $\alpha = \sqrt{n_{cts}}$; The formula of priority is designed according to a general method in reinforcement learning for achieving automatic exploitation versus exploration trade-off based on \emph{Upper Confidence Bounds}. It is motivated to focus the key search on the most promising ciphertext structures~\cite{gohr2019improving}.
    \item Pick the ciphertext structure with the highest priority score for further processing in this $j$-th iteration, denote it by $\mathcal{C}$, and its index by $idx$, $n_{\text{visit}idx} \leftarrow n_{\text{visit}idx}+1$.
    \item Run \textsc{BayesianKeySearch} Algorithm~\cite{gohr2019improving} with $\mathcal{C}$, the $r$-round neural distinguisher $\mathcal{ND}^{r}$ and its wrong key response profile $\mathcal{ND}^{r}\cdot \mu$ and $\mathcal{ND}^{r}\cdot \sigma$, $n_{cand1}$, and $n_{byit1}$ as input parameters; obtain the output, that is a list $L_{1}$ of $n_{byit1}\times n_{cand1}$ candidate values for the last subkey and their scores, \emph{i.e.}, $L_{1}=\{(g_{1i},v_{1i}):i \in \{1,\ldots,n_{byit1}\times n_{cand1}\}\}$.
    \item Find the maximum $v_{1\text{max}}$ among $v_{1i}$ in $L_{1}$, if $v_{1\text{max}} > \omega_{\text{max}idx}$, $\omega_{\text{max}idx} \leftarrow v_{1\text{max}}$.
    \item For each of recommended last subkey $g_{1i} \in L_{1}$, if the score $v_{1i}>c_{1}$,
        \begin{enumerate}
        \item Decrypt the ciphertext in $\mathcal{C}$ using the $g_{1i}$ by one round and obtain the ciphertext structures $\mathcal{C}'$ of $(1+s+r)$-round \simeck32/64.
        \item Run \textsc{BayesianKeySearch} Algorithm~\cite{gohr2019improving} with $\mathcal{C}'$ , the neural distinguisher $\mathcal{ND}^{r-1}$ and its wrong key response profile $\mathcal{ND}^{r-1}\cdot \mu$ and $\mathcal{ND}^{r-1}\cdot \sigma$, $n_{cand2}$, and $n_{byit2}$ as input parameters; obtain the output, that is a list $L_{2}$ of $n_{byit2}\times n_{cand2}$ candidate values for the last subkey and their scores, i.e., $L_{2}=\{(g_{2i},v_{2i}):i \in \{1,\ldots,n_{byit2}\times n_{cand2}\}\}$.
        \item Find the maximum $v_{2i}$ and the corresponding $g_{2i}$ in $L_{2}$, and denote them by $v_{2\text{max}}$ and $g_{2\text{max}}$.
        \item If $v_{2\text{max}}> best_{score}$, update $best_{score}\leftarrow v_{2\text{max}}$, $best_{key}\leftarrow (g_{1i},\allowbreak g_{2\text{max}})$, $best_{pos}\leftarrow idx$.
            \end{enumerate}
    \item If $best_{score}>c_{2}$, go to Step~\ref{final}.
    \end{enumerate}
\item\label{final} Make a final improvement using \textsc{VerifierSearch}~\cite{gohr2019improving} on the value of $best_{key}$ by examining whether the scores of a set of keys obtained by changing at most 2 bits on top of the incrementally updated $best_{key}$ could be improved recursively until no improvement obtained, update $best_{score}$ to the best score in the final improvement; If $best_{score} > Gbest_{score}$, update $Gbest_{score}\leftarrow best_{score}$, $Gbest_{key}\leftarrow best_{key}$.
\item Return $Gbest_{key},Gbest_{score}$.
\end{enumerate}

\end{document}